# Enhance Robustness of Image-in-Image Watermarking through Data Partitioning

Hossein Bakhshi Golestani
Department of Electrical Engineering
Sharif University of Technology
Tehran, Iran
bakhshigolestani@ee.sharif.ir

Shahrokh Ghaemmaghami
Electronics Research Institute
Sharif University of Technology
Tehran, Iran
ghaemmag@sharif.edu

*Abstract*— Vulnerability of watermarking schemes against intense signal processing attacks is generally a major concern, particularly when there are techniques to reproduce an acceptable copy of the original signal with no chance for detecting the watermark. In this paper, we propose a two-layer, data partitioning (DP) based, image in image watermarking method in the DCT domain to improve the watermark detection performance. Truncated singular value decomposition, binary wavelet decomposition and spatial scalability idea in H.264/SVC are analyzed and employed as partitioning methods. It is shown that the proposed scheme outperforms its two recent competitors in terms of both data payload and robustness to intense attacks.

## I. INTRODUCTION

It is a typical need in the digital multimedia market to embed commercial logos or trademarks as watermark in digital images, which has motivated a great deal of research focused on image-in-image watermarking in recent years [1-5]. Transform domain watermarking is often recommended to achieve greater robustness to attacks, while still complying with the human visual system (HVS), [1]. Mid-frequency components have been found to be suitable for embedding watermark, because they contain a major part of the signal components of mid-sensitivity that can still be manipulated imperceptibly. As a general comparison between popular transforms, Discrete Cosine Transform (DCT) based methods are more robust to lossy compressions, where Discrete Cosine Transform (DWT) schemes better resist the noise addition attacks [1]. In recent years, many papers have been published to solve robust watermarking problem in transform domains [2-4].

In [2], the authors have presented a DCT domain watermarking that uses and modifies the relationship between the DCT coefficients for the embedding. This approach has an acceptable performance at low compression ratios but fails in cases of coarser compression. To overcome this drawback, Lin et al. presented an improved DCT domain approach based on the concept of mathematical remainder that was able to extract the watermark appropriately at high compression rates [3]. This approach, however, deals only with JPEG compression and shows a poor performance in some other attacks.

To eliminate the drawbacks of DCT and DWT based watermarking, a new joint DWT-DCT transformation is proposed in [4]. In this scheme, a logo is embedded in mid-frequency DCT coefficients of 3-level DWT transformed of a host image. Since DWT is first applied to the host image, this scheme shows to be more robust to noise addition attacks, but fails in the strong, lossy compression attacks.

The drawbacks with such previous approaches in intense attacks and ignoring different priority of logo's parts motivated us to present a general algorithm to improve robustness of existing schemes against strong signal processing attacks. In this paper, *intense* attack means an attack with high distortion effect on cover signal e.g. JPEG compression with low quality factor or a median filtering attack with big filter size. We take the digital watermarking as a problem analogous to that in the reliable data communication, as mentioned in [6]. Accordingly, the error resilient techniques in the reliable communication could be employed in the digital watermarking. Forward Error Correction (FEC), back channel and Data Partitioning (DP) are some of these techniques used in H.264/AVC video codecs [9].

In this paper, a two-layer Data Partitioning (DP) based image-in-image watermarking is introduced to enhance robustness against intense attacks especially in JPEG compression with coarse quantization. Assume we want to embed a logo as a watermark into an image as a cover signal. Based on the idea of DP, we divide the logo into parts, where each part is embedded with a priority based strength. Selection of an appropriate partitioning method is one of major questions in this research to be answered. To analyze partitioning method requirements, three scenarios are studied. In the first one, the logo is decomposed based on truncated singular value decomposition [7], where the 2[nd] scenario breaks up the logo based on binary wavelet decomposition [8]. In the last one, the logo is divided into two parts based on the spatial scalability idea in the H.264/SVC video compression [9]. The critical part of the watermark (called *base* part) and less critical parts of the watermark (called *enhancement* part) are embedded in different DCT coefficients with a two-layer watermarking scheme.

The rest of the paper is organized as follows. In section II the proposed DP based watermarking scheme is explained in details. In section III, experimental results are presented, compared to other results, and discussed. Finally, a conclusion is given in section IV.

## II. PROPOSED ALGORITHM

This section explains the data partitioning idea for an appropriate partitioning method that leads to a two-layer watermarking scheme in the DCT domain.

### A. Data Partitioning

Data partitioning (DP) is widely used in advanced data transmission systems like H.264/AVC-SVC video coding [10,11]. In the DP, the data is grouped into some parts based on properties, priorities, sensitivities, etc., towards optimal organization, management, transmission, or storage of the data. The base part comprises the critical parts of the data (e.g. lower frequency components of an image or motion vectors in video coding), where the enhancement part is made of less critical data (e.g. higher frequency components of the image or inter coded residues in video coding) [9].

Assume the base part is better protected than the enhancement part. Since the critical part of data is better protected, the total effect of the channel errors are minimized [5]. Note that it's not necessary to have two real different channels. For example, the base and the enhancement part could be transmitted in same channel but with better and poorer FEC codes.

### B. Partitioning Method

There could be numerous techniques to decompose a binary watermark into parts based on their criticality, sensitivity, etc. These techniques, however, must meet three major requirements:

1) Partitions of a binary image must have limited luminance levels. Assume the watermark comprises K luminance levels. The embedding scheme must be designed such that these K levels are sufficiently distinguishable to minimize error probability in the detection process. If K is a large number, the distance between embedding levels becomes small and, to have an appropriate error probability, the embedding power should be increased. However, higher embedding power causes more severe degradation of the host quality, thus imposing a limit on the luminance level is an important requirement.

2) Partitioning methods must be reversible to reconstruct original watermark from its partitions exactly. If not, some errors are unavoidable, even in no attack conditions.

3) Degradations of the enhancement part should not propagate through the reconstructed watermark.

To illustrate these requirements, three partitioning schemes are analyzed.

In the First scenario, singular value decomposition (SVD) is employed to divide the watermark into two parts. Assume matrix X has R non-zero singular value where R is the rank of X. In the truncated SVD idea, the first P singular values (P<R) can reconstruct an approximation of X. This approximation could be considered as the base part and the difference between the base part and the original watermark forms the enhancement part as shown in Fig. 1.

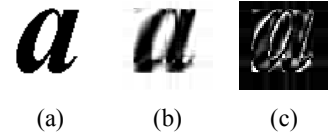

Figure 1. Truncated SVD as a partitionig method (P=5) (A) original watermark (B) base part (C) enhancement part.

This partitioning scheme does not guarantee the first requirement, while passes other criteria. Note that an error bit in the enhancement part causes exactly one error bit in the reconstructed watermark, so there is no error propagation.

In the 2$^{nd}$ scenario, a modified version of [8] is used to decompose binary watermark to four binary sub-bands. Fig. 2 shows procedure of the binary wavelet decomposition. The transform matrix used in Fig. 2 is T=[1 1 1 0; 1 0 1 1; 1 1 0 0; 0 0 1 1]. The LL band of this decomposition (Fig. 2) is used as the base part and the other 3 bands are used as the enhancement part.

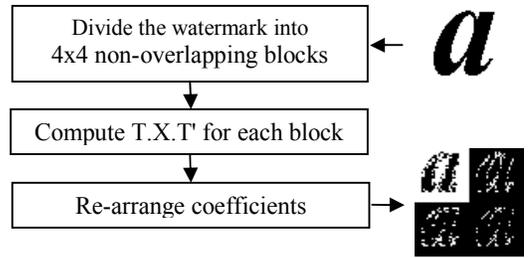

Figure 2. Binary wavelet decomposition process.

This scenario generates binary partitions and is reversible (requirements 1,2), but fails in requirement 3 for its error propagation property. Assume B is the re-arranged coefficients matrix and star mark (*) is a bit error in the B. The reconstructed watermark is obtained as follows:

$$reconstructed\ watermark = T^{-1}.B.(T')^{-1}$$

$$= \begin{bmatrix} 0 & 1 & 0 & -1 \\ 0 & -1 & 1 & 1 \\ 1 & 0 & -1 & 0 \\ -1 & 0 & 1 & 1 \end{bmatrix} \times \begin{bmatrix} . & . & . & . \\ . & . & * & . \\ . & . & . & . \\ . & . & . & . \end{bmatrix} \times \begin{bmatrix} 0 & 0 & 1 & -1 \\ 1 & -1 & 0 & 0 \\ 0 & 1 & -1 & 1 \\ -1 & 1 & 0 & 1 \end{bmatrix} \quad (1)$$

this error propagates through 6 entries of the reconstructed watermark. Changing location of the error bit in matrix B causes different error patterns in the reconstructed watermark. On average, one bit error in B results 6.25 bits error in the reconstructed watermark (error propagation), thus this scenario fails in requirement 3.

In the last scenario, the spatial scalability idea in the H.264/SVC generates two spatial resolution images, from a single source image, such that the down-sampled image is the base part and the enhancement part is the difference between the spatially interpolated base part and the original image [5] (Fig.3).

If the watermark is a binary image, then the down-sampled version of the watermark (base part) is a binary image too. But, the difference between the original watermark and the spatially interpolated base part could be "+1", "0" or "-1", therefore the enhancement part has three kinds of bits (0, +1, -1).

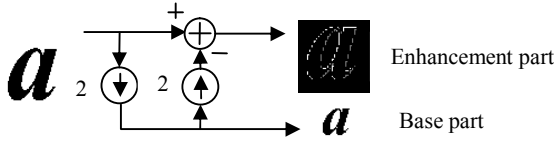

Figure 3.  Spatial scalablility as a partitioning method.

This scenario generates partitions with three luminance levels (requirement 1) and is reversible (requirement 2). An error bit in the enhancement part causes only one error bit in the reconstructed watermark, thus there is no error propagation (requirement 3). Focusing on the results indicates that spatial scalability partitioning method is better than the others because satisfies all criteria.

### C. Two-Layer Watermarking Scheme

The base part must be embedded with more power than the enhancement part, but the more embedding power results more quality degradation. To reduce the effect of coarse base part embedding, a two-layer DCT based image watermarking scheme is proposed. In this scheme, the first layer is of less capacity than that of the second layer, but spreads the blocky effect of coarse watermarking.

Without loss of generality, assume a 512x512 grayscale image as a host image. The host image is divided into 16x16 non-overlapping blocks that undergo the DCT transform, where each 16x16 DCT block is decomposed to four 8x8 sub-blocks. In this situation, the spatial relationship between DCT coefficient of 16x16 block and its 8x8 sub-blocks is given as [12]:

$$A = \frac{1}{2} P \begin{bmatrix} a & b \\ c & d \end{bmatrix} P^T \quad (2)$$

where $A$ is the DCT coefficient of 16x16 block and $a, b, c$ and $d$ are the DCT coefficients of the 8x8 sub-blocks. $P$ is a reversible 16x16 matrix that converts 16x16 DCT layer (called 16x16 layer) elements to 8x8 DCT layer (called 8x8 layer) elements and vice versa. To preserve the image quality in the watermarking process and reach the acceptable resistance against attacks, the mid-frequencies of the DCT coefficients are chosen for the embedding.

An important result which could be obtained from (2) is that $A_{1,3}$ and $A_{3,1}$ in the 16x16 layer are linear combinations of $x_{1,2}, x_{2,1}$; $x=a, b, c$ and $d$ in 8x8 layer. According to this observation, two results are deduced:

- $A_{1,3}$ and $A_{3,1}$ in the 16x16 layer are independent from $x_{1,3}, x_{3,1}$; $x=a, b, c$ and $d$ in the 8x8 layer.

- Any intense manipulation of $A_{1,3}$ and $A_{3,1}$, distributed in 8 coefficients in the 8x8 layer ($x_{1,2}, x_{2,1}$; $x=a,b,c,d$) and block effect of strong embedding is reduced.

Therefore, $A_{1,3}$ and $A_{3,1}$ are used to embed the base part, where $x_{1,3}, x_{3,1}$; $x=a,b,c,d$ are exploited to embed the enhancement part.

#### 1) Embedding and Extracting Process

In order to embed bit "1" and bit "0" in the 16x16 block, $A_{1,3}$ and $A_{3,1}$ are modified in a way that $\Delta = A_{1,3} - A_{3,1}$ is converted to nearest point in $M_1 = \{M/4, -3M/4, 5M/4 ...\}$ and $M_2 = \{-M/4, 3M/4, -5M/4, ...\}$, respectively. For example, if we want to embed "1" and $\Delta = 7M/8$, $\Delta$ is moved to $5M/4$ ($5M/4$ is the nearest point to $7M/8$ in $M_1$ set points). In other words, to embed "1" and "0", $\Delta$ is converted to the center of nearest interval that specified in Fig. 4. Embedding algorithm in 8x8 layer is the same as above while using the pattern given in Fig. 5.

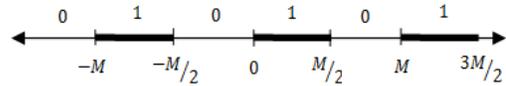

Figure 4.  Interval pattern for the base part embedding

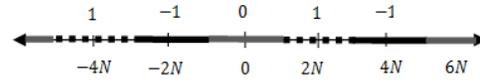

Figure 5.  Interval pattern for the enhancement part embedding.

Depending on whether $\Delta' = A'_{1,3} - A'_{3,1}$ held on the thick line or the thin line (Fig. 4), bit "1" or "0" is extracted, respectively. In this algorithm, if distortion is less than M/4, the embedded bits are correctly retrieved.

### III. SIMULATION RESULTS AND DISCUSSION

To simulate the proposed scheme, a set of 500 grayscale images of size 512x512 from Corel image database are used. We use PSNR (Peak Signal-to-Noise Ratio) and BER (Bit error rate) measures to assess the proposed method's performance and the quality degradation due to the watermark embedding.

Generally, the watermark could be either an independent and identically distributed (iid) binary sequence or a small proprietary image (e.g. a logo). If an iid binary sequence is used as a watermark, a quantitative metric is required to evaluate extraction results and determining a threshold produces another ambiguity. But, if a small meaningful image or logo is used as a watermark, human eye could easily verify the extracted result [5]. In our simulation, a binary image is employed as watermark. [This does not look to help and may rather raise some questions, so better to remove it.]

In order to analyze robustness enhancement, the normal method (without DP) and the DP based method are compared together and to two recently proposed watermarking schemes [3,4]. The spatial scalability idea in the H.264/SVC described in section III is used as the partitioning method. In the normal method, the whole of the watermark is embedded in the 8x8 layer without any partitioning. The logo size in both the proposed DP based and the normal methods is 64x64, while it is 32x32 in [3] and [4]. Thus the data payload in the DP and normal method is four times greater than those in [3] and [4]. The simulation settings and the PSNR results are shown in Table I.

TABLE I. SIMULATION SETTINGS AND AVERAGE PSNR USING DIFFRERENT EMBEDDING METHOD

| Method | Proposed DP | Normal | [3] | [4] |
|---|---|---|---|---|
| Simulation setting | M=69.1 , N=12 | M=62.2 | M=32.9 | A=19 |
| Average PSNR (dB) | 44.06 | 44.07 | 44.08 | 43.82 |

### A. JPEG Compresion Attack

To evaluate performance against the JPEG compression attack, the watermarked image was JPEG compressed with different Quality Factors. Table II shows the results.

TABLE II. COMPARISON OF METHODS IN JPEG COMPRESSION

| Quality Factor | DP | Normal | [3] | [4] |
|---|---|---|---|---|
| 60 | BER=0% | BER=0% | BER=0% | BER=21% |
| 50 | BER=1% | BER=0% | BER=0% | BER=28% |
| 40 | BER=5% | BER=5% | BER=0% | BER=34% |
| 30 | BER=9% | BER=17% | BER=13% | BER=39% |
| 20 | BER=22% | BER=38% | BER=34% | BER=42% |

To analyze the results illustrated in table II, assume the base part and the enhancement part are embedded using the DP method with powers A and B, respectively, and the whole of watermark is embedded using the normal method (no DP) with power C. To get into equal PSNRs in all methods, the embedding powers set as A > C > B. Thus:

$$R_{base} > R_{normal} > R_{enhancement}$$

where R stands for robustness. In the JPEG attack with QF=60, the degradation is below the amount to cause error in the enhancement part, so no error occur in both the DP and the normal methods. At QF=50, the degradation can cause error in the enhancement part, but not in the base part and the normal method. At QF=40, the degradation can cause error in both the enhancement part and the normal method but the base part is extracted almost perfectly. By decreasing the QF, the base part, the enhancement part, and the normal method are affected by the JPEG compression degradation, but this degradation is less effective on the base part that can help watermark extraction (see QF=20).

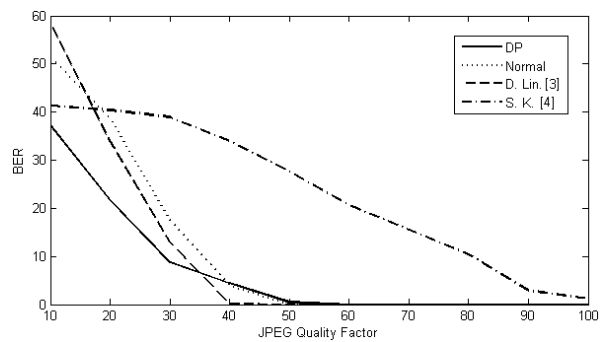

Figure 6. Comparison of BERs vs. JPEQ quality factor.

Fig. 6 shows that the DP based method is more robust in high JPEG compression ratios, as discussed earlier. As mentioned in [1], Fig. 7 illustrates that the DWT based method in [4] fails in such a lossy compression attacks.

### B. Filtering Attack

In this section, average filter, median filter, and Gaussian filter are applied to the watermarked image and the BER of the extracted watermarks are shown in tables III, IV and V, respectively.

TABLE III. BER IN AVERAGE FILTERING ATTACK

| Average Filter (size) | DP | Normal | [3] | [4] |
|---|---|---|---|---|
| 3x3 | 16.8 | 11.9 | 42.7 | 47 |
| 5x5 | 34 | 24 | 48 | 52 |
| 7x7 | 37 | 38 | 48 | 48 |
| 9x9 | 41 | 63 | 48 | 61 |

TABLE IV. BER IN MEDIAN FILTERING ATTACK

| Median Filter (size) | DP | Normal | [3] | [4] |
|---|---|---|---|---|
| 3x3 | 8.3 | 6.3 | 33 | 33 |
| 5x5 | 23.5 | 16.6 | 49 | 41 |
| 7x7 | 30 | 32 | 49 | 36 |
| 9x9 | 35 | 58 | 50 | 54 |

TABLE V. BER IN GAUSSIAN FILTERING ATTACK

| Gaussian Filter (size and sigma) | DP | Normal | [3] | [4] |
|---|---|---|---|---|
| 3x3 , 10.5 | 11.15 | 4.7 | 3.8 | 2.2 |
| 5x5 , 1.5 | 29.5 | 20 | 49 | 25.9 |
| 7x7 , 2.5 | 36.8 | 29.5 | 51 | 43.5 |
| 9x9 , 3.5 | 39.4 | 49.6 | 50 | 53 |

The last three tables illustrate that the proposed DP method is more robust to intense filtering attacks (see the last row of these tables, in particular, in cases of intense attacks). The reason for this robustness is the concept of DP as discussed earlier. In the slight attacks (first rows of the tables), the enhancement part is destroyed quickly, thus the normal method outperforms the DP in slight attacks.

### C. Noise Addition and Resize Attack

Adding Gaussian noise, salt and pepper noise have been applied to the watermarked images and the BERs of the extracted watermark are reported in tables VI and VII.

Unlike DCT based schemes, DWT based methods like [4] has a higher robustness against the noise addition attacks [1]. Tables VII and VIII demonstrate this fact but the robustness of the normal and DP based methods against this attack should be analyzed. To compare the DP based method to the normal method in the noise addition attack, assume one pepper noise dot (called pepper dot) holds in a region. This pepper dot holds in a 16x16 block, but stands in just one of its 8x8 sub-blocks (Fig. 7).

TABLE VI. BER IN GAUSSIAN NOISE ATTACK

| Gaussian noise (sigma) | DP | Normal | [3] | [4] |
|---|---|---|---|---|
| $10^{-4}$ | 0 | 0 | 0.3 | 1.5 |
| $5 \times 10^{-4}$ | 10 | 6 | 16 | 1.8 |
| $10^{-3}$ | 24 | 18 | 30 | 4.2 |
| $5 \times 10^{-3}$ | 48 | 47.7 | 50 | 17.5 |

TABLE VII. BER IN SALT & PEPPER NOISE ATTACK

| Salt and Pepper noise (%) | DP | Normal | [3] | [4] |
|---|---|---|---|---|
| 0.1 | 5.5 | 2.5 | 4.1 | 1.95 |
| 0.5 | 24 | 12.4 | 11 | 5.9 |
| 1 | 36 | 14.3 | 24 | 9.86 |
| 5 | 51 | 48 | 51.3 | 30 |

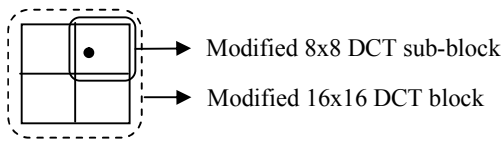

Figure 7. Effect of one pepper dot in 16x16 and 8x8 layer

This pepper dot manipulates the base part bits but alters 25% of the enhancement part bits. Since a pepper dot could affect the embedding DCT coefficient strongly, no improvement is achieved using the proposed two-layer DP based watermarking in the case of noise addition attack.

Finally, the resizing attack changes the size of the watermarked image by a certain factor (e.g. factor=0.5), and then the resized image is scaled back to its original size. Table VIII shows how the DP based method stands against the intense resizing attacks (last row in the table).

TABLE VIII. BER IN RESIZING ATTACK

| Resize (Scale factor) | DP | Normal | [3] | [4] |
|---|---|---|---|---|
| 0.8 | 0.4 | 0 | 2 | 6.8 |
| 0.6 | 3.6 | 2.5 | 2.24 | 14.5 |
| 0.4 | 14.8 | 15 | 42.5 | 26 |
| 0.2 | 36.6 | 45 | 52 | 46.7 |

IV. CONCLUSION

To improve robustness of watermarking schemes against some intense typical attacks, a data partitioning (DP) based image watermarking approach has been presented. Three criteria for a reliable partitioning method have been addressed. As a partitioning method, truncated singular value decomposition, binary wavelet decomposition, and spatial scalability idea in the H.264/SVC video coding are analyzed and discussed. The simulation results have demonstrated that the DP idea significantly improves the watermarking robustness over that of the normal methods, specifically in cases of intense attacks. Sensitivity of the DCT based two-layer watermarking to noise makes the DP based scheme ineffective in noise addition attacks.